\documentstyle[12pt]{article}
\def\etal{{\it et al.}}
\def\lsim{\:\raisebox{-0.5ex}{$\stackrel{\textstyle<}{\sim}$}\:}

\def\rp{$R_p \hspace{-1em}/\;\:$}
\def\phot{\widetilde{\gamma}}
\def\s{{\rm s}}
\def\vis{{\rm visible}}
\def\z{\widetilde{Z}}
\def\mev{\; \rm MeV}
\def\gev{\; \rm  GeV}
\begin{document}
\setcounter{page}{0}
\renewcommand{\thefootnote}{\fnsymbol{footnote}}
\thispagestyle{empty}
\vspace*{-1cm}
\begin{flushright}
CERN-TH/95--299 \\ OUTP-95-45P \\[2ex]
{\large \tt hep-ph/9511357} \\
\end{flushright}
\vskip 45pt
\begin{center}
{\Large \bf A Supersymmetric Resolution of the KARMEN Anomaly}\\
\vspace{15mm}
{\bf Debajyoti Choudhury\footnote{debchou@surya11.cern.ch}}\\ 
\vspace{10pt}
{\sl Theory Division, CERN, CH--1211 Geneva 23, Switzerland}\\
\vspace{10pt}
and \\
\vspace{10pt}
{\bf Subir Sarkar\footnote{PPARC Advanced Fellow}}\\
\vspace{10pt}
{\sl Theoretical Physics, University of Oxford, 1 Keble Road, Oxford
 OX1 3NP, UK}\\
\vspace{50pt}
{\bf ABSTRACT}
\end{center}
\begin{quotation}
We consider the hypothesis that the recently reported anomaly in the
time structure of signals in the KARMEN experiment is due to the
production of a light photino (or Zino) which decays radiatively due
to violation of $R$-parity. Such a particle is shown to be consistent
with all experimental data and with cosmological nucleosynthesis.
There are difficulties with constraints from SN~1987A but these may be
evaded if squarks are non-degenerate in mass.
\end{quotation}
\vspace{1cm}
\begin{center}
{\it Physics Letters B (in press)}
\end{center}
\vfill
\newpage
\setcounter{footnote}{0}
\renewcommand{\thefootnote}{\arabic{footnote}}
\setcounter{page}{1}
\pagestyle{plain}
\advance \parskip by 10pt

Recently, the KARMEN collaboration has reported an anomaly \cite{KAR}
in the time distribution of charged and neutral current events induced
by neutrinos emanating from
$\pi^{+}~\to~\mu^{+}\nu_{\mu}~\to~e^{+}\nu_{e}\bar{\nu_{\mu}}$ decays
at rest. After all the $\pi^+$s have decayed in the beam stop, one
expects to see an exponential distribution characterized by the muon
lifetime, but the data reveal an additional `bump' containing
$125\pm23$ events~\cite{KAReps} which arrive 3.6~$\mu\s$ after
beam-on-target. The KARMEN collaboration have suggested that this may
be due to a new massive slow-moving ($\beta_{x}\sim0.02$) particle
$x$, which is produced through $\pi^{+}\to\mu^{+}x$. The mean flight
path between the beam stop and detector is 17.5~m (including over 7~m
of steel shielding) and no anomaly is observed in the visible energy
deposition in the detector, which is consistent with that of ordinary
neutrino events which have $\langle\,E_{\vis}\,\rangle\sim11-35$
MeV. Thus this hypothetical particle $x$ must be neutral and weakly
interacting with a mass $m_{x}=33.9$~MeV, and decay (producing
electromagnetic energy) with a lifetime $\tau_{x}$ exceeding
0.3~$\mu\s$.

The possibility that $x$ is a neutrino has been examined in some
detail~\cite{bps}. While $x$ {\em cannot} be a doublet neutrino, in
particular the $\nu_{\tau}$, it can consistently be interpreted as an
isosinglet (sterile) neutrino produced through its mixing with the
$\nu_{\mu}$ and decaying through its mixing with all three doublet
neutrinos. The dominant (visible) decay modes,
$x~\to~e^{+}e^{-}\nu,~\nu\gamma$, proceed faster than usual due to the
absence of GIM suppression, with a combined width~\cite{bps}
\begin{equation}
 \Gamma_{\vis} = K \left[920 |U_{ex}|^2 + 210 |U_{\mu x}|^2 
                         + 210 |U_{\tau x}|^2\right]\;{\s}^{-1},
\end{equation}
where $K=1~(2)$ for Dirac (Majorana) $x$. The KARMEN event
rate~\cite{KAR} determines the product
\begin{equation}
 B\ (\pi^{+}\to\mu^{+}x)\ \Gamma_{\vis} \simeq 3\times10^{-11} {\s}^{-1}
\end{equation}
with the branching ratio
$B=0.0285 |U_{\mu x}|^2$~\cite{bps}. An ingenious experiment at
PSI~\cite{daum} has recently imposed the stringent upper limit
\begin{equation}
 B\ (\pi^{+}\to\mu^{+}x) < 2.6 \times 10^{-8}\ (95\%\ {\rm c.l.}),
\end{equation}
implying that $\Gamma_{\vis}>1.1\times10^{-3}$. This can still be
satisfied, given present experimental limits, for a reasonably large
region in the $|U_{e{x}}|^2-|U_{\mu{x}}|^2-|U_{\tau{x}}|^2$ parameter
space. The favoured solutions require $|U_{\tau{x}}|^2$ to dominate
$\Gamma_{\vis}$ since this allows a short lifetime
$\tau_{x}\sim0.001-150~\s$, whereas when $|U_{e{x}}|^2$ or
$|U_{\mu{x}}|^2$ dominate, the lifetime is
$\tau_{x}\sim150-300~\s$~\cite{bps}. Such short lifetimes are
necessary to evade constraints coming from cosmological and
astrophysical considerations, viz. that $x$ neutrinos produced
(through matter-enhanced mixing) in the early universe should decay
before the nucleosynthesis era and that $x$ particles produced (by
nucleon bremsstrahlung) in the core of SN~1987A should decay within
the core itself.

In this Letter, we investigate a different possibility, viz. that $x$
is the lightest neutralino in the minimal supersymmetric standard
model (MSSM)~\cite{mssm}. We shall assume that this is the lightest
supersymmetric particle (LSP), which is either the photino ($\phot$)
or the Zino ($\z$). We show that this hypothesis too is consistent
with all experimental data and fares no worse with regard to the
cosmological and astrophysical constraints.

It is obvious that a straightforward supersymmetrization of the SM
Lagrangian does not lead to the process
\begin{equation}
    \pi^+ \to \mu^+ + \phot \ (\z)\ ,
         \label{prodn}
\end{equation}
since this does not conserve lepton number. However, the most general
gauge-invariant superpotential contains, apart from the usual terms,
pieces such as \cite{superpot}
\begin{equation}
    {\cal W}_{\not R} =  \lambda_{ijk} L_i L_j E^c_k
                        +  \lambda'_{ijk} L_i Q_j D^c_k
                        +  \lambda''_{ijk} U^c_i D^c_j D^c_k \ ,
\label{R-parity}
\end{equation}
where $L_i$ and $Q_i $ are the $SU(2)$-doublet lepton and quark
superfields, and $E^c_i, U^c_i, D^c_i$ are the singlet
superfields. Clearly $\lambda_{ijk}$ is antisymmetric under the
interchange of the first two indices, while $\lambda''_{ijk}$ is
antisymmetric under the interchange of the last two. To avoid the
potential embarrassment that the introduction of such terms may cause
(e.g. the $\lambda$ and $\lambda'$ couplings violate lepton number
conservation, while $\lambda''$ couplings violate baryon number), a
discrete symmetry termed $R$-parity is introduced. Representable as
$R=(-1)^{3B+L+2S}$~\cite{rpardef}, where $B,L,S$ are the baryon
number, lepton number and the intrinsic spin of the field
respectively, this symmetry rules out each of the terms in
eq.(\ref{R-parity}), with the additional important consequence that
the LSP is required to be absolutely stable.

Although an exact $R$-parity may be phenomenologically desirable, it
is not essential from a theoretical point of view. Thus we may allow
for $R$-parity--violating (\rp) couplings as long as they are not
inconsistent with experimental data~\cite{limits,sponrpar}; for
example, rapid proton decay can be simply prevented by requiring that
all the $\lambda''_{ijk}$ be zero. This is in fact well motivated
theoretically~\cite{rossiban}; moreover such a scenario has
interesting cosmological implications, e.g. for
baryogenesis~\cite{baryo}. It has been argued that the presence of
other \rp\ terms may wash out the baryon asymmetry of the
universe~\cite{camp}; however this is model-dependent and can be
evaded, for example through lepton mass effects which allow a baryon
asymmetry to be regenerated at the electroweak scale through sphaleron
processes if there is a primordial flavour-dependent lepton
asymmetry~\cite{krs-dr}. Moreover, there are other ways in which a
primordial asymmetry may be protected or regenerated~\cite{altbaryo}.

The presence of \rp\ couplings makes it possible for the LSP to decay,
thus allowing the $\phot~(\z)$ to deposit visible energy in the KARMEN
detector in our model. The preferred mode of decay is, of course,
determined by both the spectrum of the theory and the relative sizes
of the possible \rp\ couplings. For the process in eq.(\ref{prodn}) to
operate, $\lambda'_{211}$ must be non-zero. The strongest constraint
on this coupling comes from charged-current universality
\cite{barger}:
\begin{equation} \displaystyle
    \lambda'_{211} \leq 0.09\:
    \left(\frac{m_{\tilde q} }{100\gev}\right)\ .
      \label{const1}
\end{equation}
Bounds on the other couplings are obtained from both low-energy
measurements such as meson decays~\cite{barger,d_tau} and limits on
$\Delta\,L=2$ operators~\cite{grt}, as well as the LEP measurements at
the $Z$--peak \cite{Zpeak}. Except for the bound on $\lambda'_{111}$
(which is required to be $\lsim 10^{-4}$~\cite{hkk}), the other bounds
are similar to that shown in eq.(\ref{const1}) or weaker. Constraints
on products of couplings can however be obtained from the data on
flavour-changing neutral-current (FCNC) processes, and these are much
more severe~\cite{dc_pr}.

We can now consider two possibilities :
\begin{enumerate}
 \item $\lambda'_{211}$ is the {\em only} \rp\  coupling in the theory, so
        the neutralino may decay only radiatively.
 \item \rp\  couplings other than $\lambda'_{211}$ are non-zero, so that
        the neutralino may decay (at tree level) into $e^+e^-\nu$ etc.
\end{enumerate}
Examination of FCNC processes demonstrate that bounds from such data
on the products of \rp\ couplings are least restrictive when only one
flavour is nonconserved~\cite{dc_pr}. Hence, for the second
possibility above, this implies that the second non-zero \rp\ coupling
should be $\lambda_{211}$. Since electron flavour is then a good
quantum number, bounds from $\mu{\to}e\gamma$ and $\mu\to3e$ are
inoperative while constraints from other muon-number-violating
processes are comparatively weaker. Thus the situation is analogous to
that in ref.\cite{bps}, albeit with relatively more freedom in the
parameter space.\footnote{Such a phenomenology also arises if there is
mixing between the neutrinos with the gauginos/Higgsinos which gives
rise to a non-zero sneutrino vev (as in spontaneous \rp\
models~\cite{sponrpar}); the neutralino decay to $e^+e^-\nu$ may then
explain the KARMEN anomaly but this requires the Higgs mixing term
$\mu\,H_1\,H_2$ in the superpotential to be bounded by $\mu\leq30$~MeV
\cite{marek} which would further exacerbate the well-known $\mu$
hierarchy problem~\cite{gm88}. This scenario also implies a massive
$\nu_{\tau}$ which is severely constrained by many cosmological and
astrophysical arguments, in particular those concerning
SN1987A~\cite{gr95}.} We shall not discuss this any further.

Returning to the first possibility, we shall limit our discussion to
the photino case; for the $\z$, one obtains analogous results. The
dominant decay mode of the photino is the radiative one,
\begin{equation}
   \phot \to \nu_{\mu} + \gamma \ ,
\label{decay}
\end{equation}
proceeding through a one-loop diagram involving the $d$-quark and the
$\tilde d_{L,R}$-squarks. The relevant $\pi$-decay (see eq.\ref{prodn})
branching fraction is then
\begin{equation} 
\begin{array}{rcl}
\displaystyle
   {\cal P} \equiv \frac{\Gamma(\pi^+ \rightarrow \mu^+ \phot)}
     {\Gamma(\pi^+ \rightarrow \mu^+ \nu_\mu)}
& = & \displaystyle \frac{8\pi}{9 \alpha}
   \left(\frac{\lambda'_{211} 
   \sin^2 \theta_W m_\pi^3}{m_u + m_d} \right)^2 \:
   \frac{m_\pi^2 - m_\mu^2 - m_{\phot}^2}{ m_\mu^2 (m_\pi^2 - m_\mu^2)^2} \\
& & \displaystyle \hspace*{1em} {\cal C}\negthinspace\negthinspace
         \left(1, \frac{m_{\phot}^2}{m_\pi^2},
                  \frac{m_\mu^2}{m_\pi^2} \right) \: m_W^4 
    \left[ \frac{1}{m^2_{\tilde d R} } 
          + \frac{2}{m^2_{\tilde u L} }  
          -  \frac{6}{m^2_{\tilde\mu L} } \right]^2
\label{prod_cs}
\end{array}
\end{equation}
where ${\cal C}(a,b,c) \equiv \left[ (a - b - c)^2 - 4 b
c\right]^{1/2}$ is the usual Callan function. Henceforth, we shall
assume a common sfermion mass
$m_{\rm sf} \equiv m_{\tilde u_L} = m_{\tilde d_L}
                 = m_{\tilde d_R} = m_{\tilde \mu L}$. 
Using the constituent quark masses, we have
\begin{equation} \displaystyle
    {\cal P} \simeq 8.9 \times 10^{-4} \;\lambda^{\prime 2}_{211}
     \left( \frac{100\gev}{m_{\rm sf}} \right)^4 .
\label{prod-num}
\end{equation}
On the other hand, the photino decay width is
\begin{equation} \displaystyle
    \Gamma \equiv
    \Gamma(\phot \rightarrow \nu_\mu \gamma)
    = \frac{\alpha^2 \lambda^{\prime 2}_{211} }{2592 \pi^3}\;
    \frac{m_{\phot}^3}{m^2_{\rm sf}} \;\:
    f\negthinspace\negthinspace\left(\frac{m_d^2}{m^2_{\rm sf}}\right)\ ,
\label{decaywidth}
\end{equation}
where
\begin{equation}
   f(x) = x \left[ 1 +\frac{3 - 4 x + x^2 + 2 \ln x}{(1 - x)^3}\right]^2 .
\end{equation}
In fig.~1, we show the correlation between ${\cal P}$ and lifetime
$\tau\equiv\Gamma^{-1}$ required to reproduce the experimental result,
with the part consistent with our hypothesis highlighted. The latter
information is also presented in fig.~2 in the form of a dark band
in the $m_{\rm sf}$--$\lambda'_{211}$ plane. For parameter values in
this region, the quantities ${\cal P}$ and $\Gamma$ would obey the
correlation of fig.~1. We now proceed to examine the consistency of
this hypothesis {\em vis a vis} cosmological and astrophysical bounds.

\begin{figure}[tbh] \label{correl}
\vskip 5.3in\relax\noindent\hskip -0.8in\relax{\includegraphics{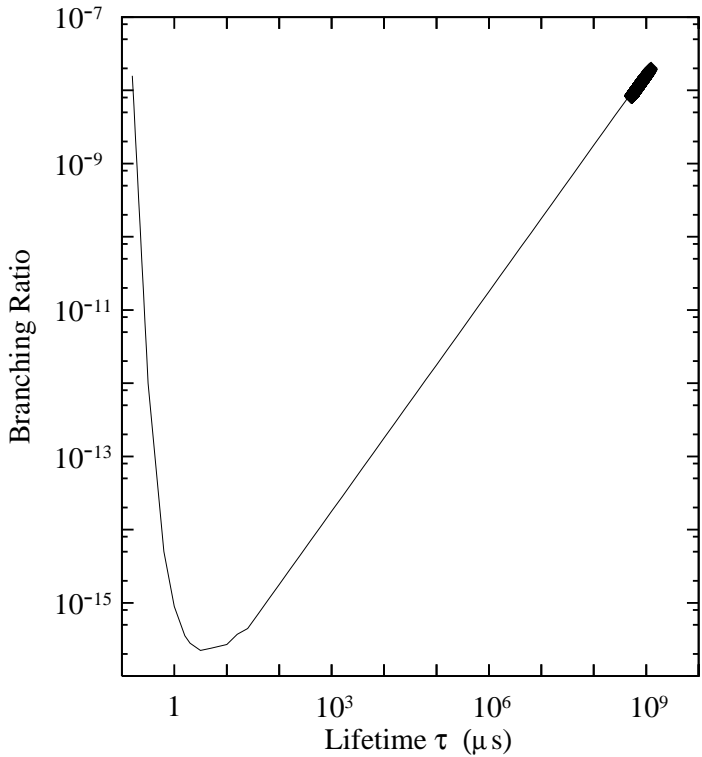}}
\vspace{-18ex}
\caption{The correlation between the mean lifetime and the production
branching ratio required to explain the KARMEN anomaly. The part
consistent with the decaying photino hypothesis (assuming degenerate
squarks) is emphasized in bold.}
\end{figure} 

\begin{figure}[tbh] \label{param}
\vskip 5.3in\relax\noindent\hskip -0.8in\relax{\includegraphics{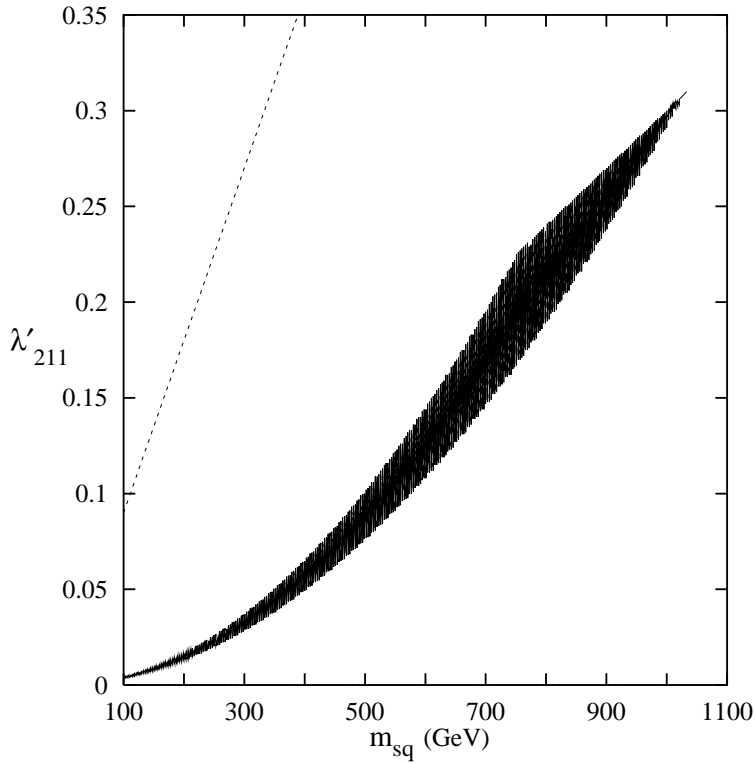}}
\vspace{-18ex}
\caption{The region in parameter space (dark band) which admits the
decaying photino solution to the KARMEN anomaly. The dashed line shows
the experimental upper bound ($ 1 \sigma$) on the \rp\ coupling as a
function of the (assumed common) squark mass.}
\end{figure} 

A potential problem with the production of a massive unstable particle
in the early universe is that this can disrupt the standard cosmology
which is in good agreement with observations~\cite{kt}. For example,
the mass density of a non-relativistic particle can speed up the
expansion rate during nucleosynthesis, while the electromagnetic
energy generated through its subsequent decays can dilute the
nucleon-to-photon ratio, resulting in the synthesis of too much
$^{4}$He. Decays that create high energy photons can also alter the
elemental abundances through photofission processes or distort the
blackbody spectrum of the relic 2.73 K radiation. Such considerations
enable stringent bounds to be placed on the relic abundance of the
decaying particle as a function of its lifetime~\cite{cosmo}. Using
the standard freeze-out formalism~\cite{kt}, we find the relic
abundance of the hypothetical photino to be
\begin{equation}
   m_{\phot} \left(\frac{n_{\phot}}{n_{\gamma}}\right) \sim 
    2.4 \times 10^{-3}\gev \left(\frac{m_{\phot}}{34 \mev}\right)^{-2}
     \left(\frac{m_{\tilde{q},\tilde{l}}}{100\gev}\right)^{4}\ .
\label{cosmoabund}
\end{equation}
This is rather high\footnote{The relic abundance of a singlet
neutrino~\cite{bps} is also high since the (matter-enhanced)
oscillation processes that create it go out of equilibrium at a
temperature of ${\cal O}(\gev)$, when it is still
relativistic. Although its number density relative to doublet
neutrinos is thus diluted by a factor of $\sim10$ during the
subsequent quark--hadron transition, its energy density during
nucleosynthesis is very large since by this time the singlet neutrinos
have turned non-relativistic, thus
$m_{\nu_\s}n_{\nu_\s}/n_{\gamma}\sim
1.9\times10^{-3}\gev\,(m_{\nu_\s}/34\mev)$.}  because the
self-annihilation cross-section is (s-wave) suppressed for a Majorana
particle~\cite{photann}. We see from ref.~\cite{cosmo} that the decay
lifetime is then required to be less than a few hundred seconds in
order that the synthesised $^{4}$He mass fraction not exceed the
conservative upper limit of $25\%$. This constraint is satisfied for
$\lambda'_{211}$ close to its highest permissible value consistent
with the KARMEN event rate.

Light photinos can also be produced through nucleon bremsstrahlung and
$e^{+}e^{-}$ annihilation in supernovae such as
SN~1987A~\cite{snreview}.  If the squark and selectron masses are
comparable to $m_{W}$ then the photinos are trapped in the dense
core by photino-nucleon and photino--electron elastic scatterings and
diffuse out to be emitted from a `photinosphere' with a thermal
spectrum, like an additional species of neutrino. The temperature of
the photinosphere increases, thus increasing the photino luminosity,
as the elastic scattering cross-section is decreased by increasing the
squark/slepton mass. Thus an {\em upper} bound on the latter follows
by considerations of the total energy loss permitted. However, as the
slepton/squark mass is further increased, photino interactions with
matter eventually become so weak that they begin to escape freely. At
this point the photino luminosity peaks and then begins to decrease,
and so can be made phenomenologically acceptable for a sufficiently
large slepton/squark mass. It has been argued that consistency with
observations of SN~1987A is not possible for any squark mass which is
allowed by experiment or `naturalness' arguments, and that a stable
light photino is therefore ruled out
altogether~\cite{snphotino}.\footnote{This conclusion may be evaded
only if selectrons are significantly lighter than
squarks~\protect\cite{lau}.}  To evade this constraint one must invoke
\rp\ violation to make the photino unstable as in the present case,
but other constraints then come into play.

If the lifetime is longer than $\sim10^{3}\s$, the escaping photinos
would decay outside the supernova. This would have resulted in a
gamma-ray flash which was not seen from SN~1987A by satellite-based
detectors~\cite{satellite}. For shorter lifetimes, the decays would
have occurred within the progenitor, and the decay photons would have
been thermalized leading to distortions of the lightcurve. Since the
observed lightcurve of SN~1987A appears to be well understood in terms
of energy input from $^{56}$Co decay (but see ref.\cite{hatsuda}),
this would appear to rule out such decays unless the lifetime is so
short than decays occur within the core~\cite{cowsik}. This
requirement, viewed in conjunction with fig.~2, apparently excludes
the hypothesis under consideration, but there are possible
loopholes. For example, we have assumed throughout that all three
relevant squarks as well as the smuon are mass-degenerate. While the
excellent agreement found at LEP between the $\rho$-parameter and its
SM value demands that $m_{\tilde u_L}\approx\,m_{\tilde d_L}$, the
mass of the $\tilde d_R$ need not satisfy this constraint. In fact
only $\tilde d_{L,R}$ contributes to the photino lifetime. On the
other hand, the branching ratio receives contributions from both
$\tilde u_L$ and $\tilde d_R$; for identical squark masses, the
$\tilde u_L$ contribution is {\em larger}. Assuming that $m_{\tilde
d_R}<m_{\tilde u_L}\approx\,m_{\tilde d_L}$ thus allows us to
significantly lower the photino lifetime while still being in
agreement with the correlation of fig.~1 and consistent with the weak
lower limit to the photino lifetime from the VENUS
experiment~\cite{venus}. This can also be effected (to a greater
degree) if the smuon mass were to be somewhat larger than the common
squark mass. One should also reconsider the production rate of
photinos with a mass as large as 34 MeV, since this is of the same
order as the core temperature.

To summarize, the KARMEN anomaly can be interpreted as due to the
production and decay of photinos in a supersymmetric theory with
$R$-parity violation. Just {\em one} non-zero \rp\ coupling suffices
for this purpose, while non-zero values for more than one \rp\
coupling makes the phenomenology similar to that of an unstable
sterile neutrino. The lifetime for radiative decays is consistent with
bounds from cosmological nucleosynthesis. The only problematic
constraint comes from the light curve of SN~1987A, but this may
possibly be evaded by assuming a hierarchy in squark masses. An
experimental test of this hypothesis would be look for the
monoenergetic decay photons with $E_{\gamma}\simeq17\,\mev$ which
distinguishes this from the singlet neutrino scenario (as well as the
\rp\ scenario with more than one \rp\ coupling) where the decay
electrons are expected~\cite{bps} to have a characteristic broader
spread in energy.

{\bf Acknowledgements:} DC wishes to thank G.~Bhattacharyya,
M.~Neubert and F.~Vissani for discussions and SS would like to thank
H.~Dreiner, A.~Pilaftsis, R.J.N.~Phillips and B.~Seligmann. SS is a
PPARC Advanced Fellow and acknowledges support from the EC Theoretical
Astroparticle Network.

\end{document}